\documentclass{article}

\usepackage{PRIMEarxiv}

\usepackage[utf8]{inputenc} 
\usepackage[T1]{fontenc}    
\usepackage{hyperref}       
\usepackage{url}            
\usepackage{booktabs}       
\usepackage{amsfonts}       
\usepackage{nicefrac}       
\usepackage{microtype}      
\usepackage{lipsum}
\usepackage{fancyhdr}       
\usepackage{graphicx}       
\graphicspath{{media/}}     

\usepackage{algorithm}
\usepackage{multirow}
\usepackage{multicol}
\usepackage{algpseudocode}
\usepackage{subcaption}

\usepackage{amsthm}

\newtheorem{theorem}{Theorem}
\newtheorem{lemma}[theorem]{Lemma} 
\title{Community Detection using Fortunato's Performance Measure
}

\author{
  Srushti Thakar \\
  School of Engineering and Applied Science,\\
  Ahmedabad University\\
  \texttt{srushti.t@ahduni.edu.in} \\
   \And
  Amit A. Nanavati \\
 School of Engineering and Applied Science,\\
  Ahmedabad University\\
  \texttt{amit.nanavati@ahduni.edu.in} \\
}

\begin{document}
\maketitle

\begin{abstract}
In his paper on Community Detection~\cite{fortunato2010community}, Fortunato introduced a quality function called {\it performance} to assess the goodness of a graph partition. This measure counts the number of correctly ``interpreted” pairs of vertices, i. e. two vertices belonging to the same community and connected by an edge, or two vertices belonging to different communities and not connected by an edge. In this paper, we explore Fortunato's performance measure ({\it fp} measure) for detecting communities in unweighted, undirected networks. First, we give a greedy algorithm {\it fpGreed} that tries to optimise the {\it fp} measure by working iteratively at two-levels, vertex-level and community-level.  At the vertex level, a vertex joins a community only if the {\it fp} value improves. Once this is done, an initial set of communities are obtained. At the next stage, two communities merge only if the {\it fp} measure improves. Once there are no further improvements to be made, the algorithm switches back to the vertex level and so on. {\it fpGreed} terminates when there are no changes to any community. We then present a faster heuristic algorithm {\it fastFp} more suitable for running on large datasets. We present the quality of the communities and the time it takes to compute them on several well-known datasets. For some of the large datasets, such as {\it youtube} and {\it livejournal}, we find that Algorithm {\it fastFP} performs really well, both in terms of the time and the quality of the solution obtained.
\end{abstract}


\section{Introduction}

Community detection in networks has always been an important problem, with applications ranging from social and communication networks to biological systems. 
In their recent paper, ``20 years of network community detection", the authors state that {\it the fundamental technical challenge in the analysis of network data is the automated discovery of
communities}~\cite{fortunato202220}. A generally accepted principle is that links within communities should be relatively frequent, while those between communities should be relatively rare~\cite{traag2011narrow}.

In order to distinguish between ``good” and ``bad” partitions of a network into communities, it would be useful to have a quantitative criterion to assess the goodness of a graph partition. A quality function is a function that assigns a number to each partition of a graph~\cite{fortunato2010community}. Modularity~\cite{newman2004finding} is one such quality function. It is based on the idea that a random graph is not expected to have a cluster structure, and therefore requires a suitable null model for its calculation. Apart from being a quality function, modularity optimization is itself a popular method for community detection and has been studied extensively~\cite{clauset2004finding,blondel2008fast,traag2019louvain,fortunato2007resolution}.

The same paper~\cite{fortunato2010community} introduced another quality function {\it performance} (henceforth, {\it fp}, for ``Fortunato's performance"), that counts the number of correctly ``interpreted” pairs of vertices, i. e. two vertices belonging to the same community and connected by an edge, or two vertices belonging to different communities and not connected by an edge. For a partition ${\mathcal P}$, the {\it fp} measure is calculated as:
\begin{equation}
{\textit fp}({\mathcal P}) = { |\{(i,j)\in E, C_i = C_j\}| +|\{(i,j)\not\in E, C_i\not= C_j\} |\over {n(n-1)/2}}
\end{equation}
$0\leq {\textit fp}({\mathcal P})\leq 1.$ This is a well-known quality measure and has been implemented in libraries such as Networkx~\cite{networkx} for measuring the ``goodness" of a network partition. The {\it fp} measure explicitly captures the intuitive notion that the more the number of intra-community edges, and the less the number of inter-community edges, the better. It does not account for null models based on random-graphs. While modularity (also a quality measure) has been explored extensively for community detection, to the best of our knowledge and somewhat surprisingly, {\it fp} has not. 
In this paper, we explore the {\it fp} measure as an optimisation metric for community detection in networks. 

Since modularity is a very popular metric, it has been studied extensively, and some of its limitations are now well-understood. Modularity contains an
intrinsic scale that depends on the total number of links in the
network~\cite{fortunato2007resolution}. Modules that are smaller than this scale (resolution limit) may not be resolved, even in the extreme case where they are complete graphs connected by single bridges. The {\it fp} measure is simple, intuitive, and does not suffer from resolution-limit issues.




{\it Our Contribution.} 
We discuss the suitability of the {\it fp} measure as a quality metric and highlight the advantages of using it in community detection. We present two algorithms: Algorithm {\it fpGreed} that greedily merges nodes into communities to maximise the {\it fp} measure. Algorithm {\it fastFP} employs a faster heuristic, suitable for running on large networks. We experimented with these two algorithms on various real-life datasets to find the {\it fp} values and the time it took to obtain them.

In Section~\ref{sec:related}, we discuss related work. Section~\ref{sec:fp-opt} details Algorithm {\it fpGreed} which greedily optimises the {\it fp} measure at the vertex-level and the community-level. We show the results obtained from running Algorithm {\it fpGreed} on a few well-known networks. Section~\ref{sec:fast} discusses Algorithm {\it fastFp}, a faster heuristic, especially suitable for large networks. Section~\ref{sec:expt} tabulates the comparison of running both the algorithms on various datasets. Section~\ref{sec:conclude} discusses the limitations of this paper and potential future work.


\section{Why optimise {\it fp}?}
\label{sec:related}

\begin{figure}[htbp]
    \centering
    \begin{subfigure}{0.45\linewidth}
        \centering
        \includegraphics[width=0.7\linewidth]{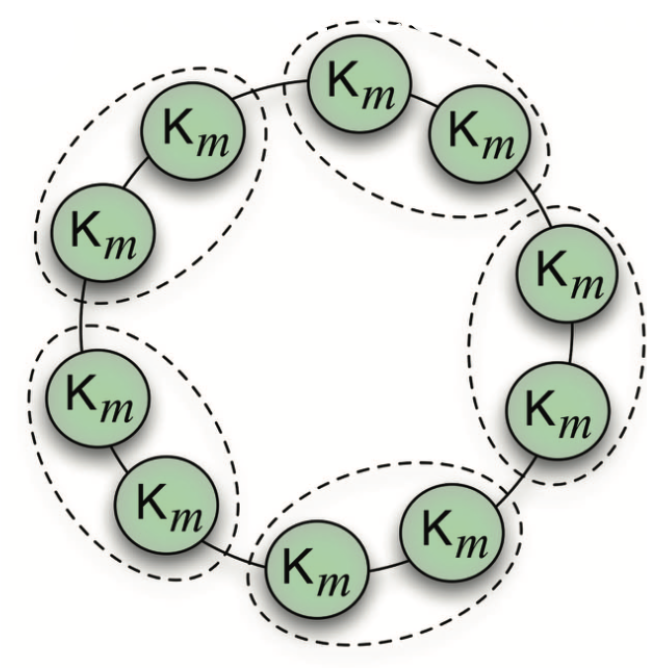}
        \caption{}
        \label{fig:sub1}
    \end{subfigure}
    \hspace{\fill}
    \begin{subfigure}{0.45\linewidth}
        \centering
        \includegraphics[width=0.7\linewidth]{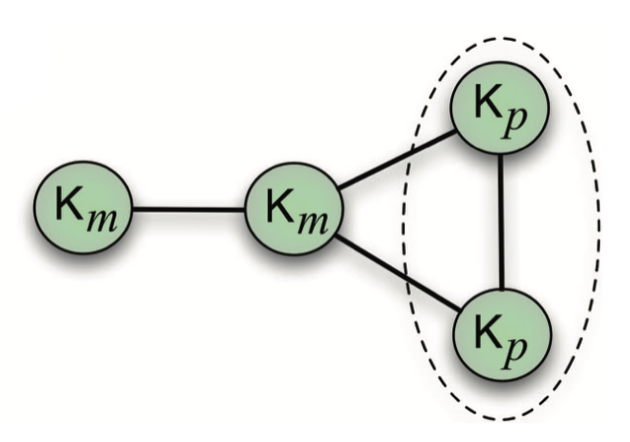}
        \caption{}
        \label{fig:sub2}
    \end{subfigure}
      \caption{Two examples from~\cite{fortunato2007resolution} to show that modularity optimisation can lead to counter-intuitive results. In both the figures, modularity optimization leads to a partition where the cliques are combined into groups of two or more (represented by dotted lines). The intuitive and correct answer in each case is for each clique to be a separate community. Optimising the {\it fp} measure gives the intuitive answer (Table~\ref{tab:res}).}
    \label{fig:main}
\end{figure}

\begin{table}[h]
    \centering
    \caption{Resolution Limit Cases (a) and (b) with m=20, p=5~\cite{fortunato2007resolution}. Unlike Modularity optimisation, {\it fp} optimisation yields the expected number of modules(communities).}
    \label{tab:merged_results_full}    \begin{tabular}{|c|ccc|ccc|}
        \hline
        \multirow{1}{*}{} & \multicolumn{3}{c|}{\textbf{modularity optimization}} & \multicolumn{3}{c|}{\textbf{fp optimization}} \\
        \cline{2-7}
         & \textbf{fp} & \textbf{mod} & \textbf{\#modules} & \textbf{fp} & \textbf{mod} & \textbf{\#modules} \\
        \hline
        (a) & 0.9706  & 0.8871 & 16 & 0.9973 & 0.8758 & 30 \\
        (b) &0.9780  & 0.5426 & 3  & 0.9967 & 0.5416 & 4 \\
        \hline
    \end{tabular}
    \label{tab:res}
\end{table}

{\it Modularity.}
Most community detection algorithms focus on the modularity optimization provided by Newman and Girvan ~\cite{newman2004finding}. Modularity is a popular quality function that compares the actual observed intra-edge density with that expected under a random null model and is defined as:
\[
Q = \frac{1}{2m} \sum_{i,j} \left( A_{ij} - \gamma \frac{k_i k_j}{2m} \right) \delta(c_i, c_j)
\]
The term $k_ik_j\over 2m$ is the expected number of edges between nodes $i$ and $j$
in a random network given their degrees.
However, modularity optimization for community detection suffers from resolution limitations, explaining its inability to identify small communities irrespective of their structural significance in the network. This limitation was analysed by Fortunato and Barthélemy ~\cite{fortunato2007resolution}, who demonstrated that modularity tends to merge smaller but meaningful communities. The authors present two examples (Figure ~\ref{fig:main}), to explain this limitation.  Table~\ref{tab:res} presents the results of modularity optimisation and {\it fp} measure optimisation. The latter is able to correctly identify the number of communities. 

{\it Potts Model Variants.}
To overcome the resolution-limit problem, the authors of~\cite{kumpula2007limited} identify a tunable parameter $\gamma$. By adjusting $\gamma$, it becomes possible to detect smaller communities. The Constant Potts Model (CPM) introduced in~\cite{traag2011narrow} tries to maximize the number of internal edges while at the same time keeping relatively small communities. The parameter $\gamma$ balances these two imperatives. In fact, the parameter $\gamma$ acts as the inner and outer edge density threshold. By setting $\gamma$, one can control the type of communities that can be detected.
On the other hand, the {\it fp} measure does not require any parameters. Its formulation naturally prevents trivial solutions such as merging all nodes into a single community, which is an issue with some other quality metrics such as coverage~\cite{fortunato2010community}.

{\it Label Propagation.}
The label propagation method~\cite{raghavan2007near}, is computationally fast and efficient. The community detection iteratively assigns labels to nodes based on majority local voting. Despite their
high speed and scalability, such methods often lack stability, as they may result in different partitions based on initializations or tie-breaking strategies~\cite{hou2017semi}. 


\section{Optimising the {\it fp} measure for Community Detection}
\label{sec:fp-opt}
\newcommand{\LineComment}[1]{\State \hspace{\algorithmicindent}{// #1}}
\algrenewcommand\algorithmicrequire{\textbf{Input:}}
\algrenewcommand\algorithmicensure{\textbf{Output:}}
\renewcommand{\thealgorithm}{}

We now present Algorithm {\it fpGreed}, which take the graph as input and outputs communities.
The idea is to first merge the nodes greedily in a way that increases the {\it fp} value, and when this is no longer possible, then merge communities greedily. Then back to nodes. The algorithm terminates when no changes happen at both levels.
\begin{algorithm}
\caption{{\it fpGreed}}
\begin{algorithmic}[1]

\Require Graph $G = (V, E)$
\Ensure A set of communities $\mathcal{C} = \{C_1, C_2, \dots, C_k\}$

\State // Initialize each node $u \in V$ as its own community
\State $\mathcal{C} \gets \{ u \in V: \{u\} \}$
\State $change \gets 1$

\While{true}
    \If{$change == 1$}
        \State $change \gets 0$
        \State {// Node-level merging}
        \ForAll{node $u \in V$}
            \State $C_u \gets$ current community of $u$
            \State {// store {\it fp} of current communities}
            \State $best\_gain \gets$ ${\it fp}(\mathcal{C})$
            \State $best\_comm \gets C_u$
            \State $NeighComms \gets$ communities containing 
            \Statex \hspace{.71 in} neighbours of $u$
            \ForAll{community $C_v \in NeighComms$}
                \State {// performance if $u$ is moved from $C_u$ to $C_v$}
                \State {// let the modified set of communities be $\mathcal{C'}$}
                \State $gain \gets$ ${\it fp}\mathcal{(C')}$
                \If{$gain > best\_gain$}
                    \State $best\_gain \gets gain$
                    \State $best\_comm \gets C_v$
                \EndIf
            \EndFor
            \If{$best\_comm \ne C_u$}
                \State move $u$ to $best\_comm$
                \State $change \gets 1$
            \EndIf
        \EndFor
    \EndIf

    \If{$change == 1$}
        \State $change \gets 0$
        \State {// Community-level merging}
        \ForAll{community $C_i \in \mathcal{C}$}
            \State {// store {\it fp} of current communities}
            \State $best\_gain \gets$ {\it fp}$(\mathcal{C})$
            \State $best\_comm \gets C_i$
            \State $NeighComms \gets$ communities adjacent to $C_i$
            \ForAll{community $C_j \in NeighComms$}
                \State $gain \gets$ performance if $C_i$ and $C_j$ are merged
                \If{$gain > best\_gain$}
                    \State $best\_gain \gets gain$
                    \State $best\_comm \gets C_j$
                \EndIf
            \EndFor
            \If{$best\_comm \ne C_i$}
                \State merge $C_i$ with $best\_comm$
                \State update $\mathcal{C}$
                \State $change \gets 1$
            \EndIf
        \EndFor
    \Else
        \State \textbf{break}
    \EndIf
\EndWhile
\State \textbf{return} $\mathcal{C}$
\end{algorithmic}
\end{algorithm}
\begin{figure}[h]
  \centering
  \includegraphics[width=0.5\linewidth]{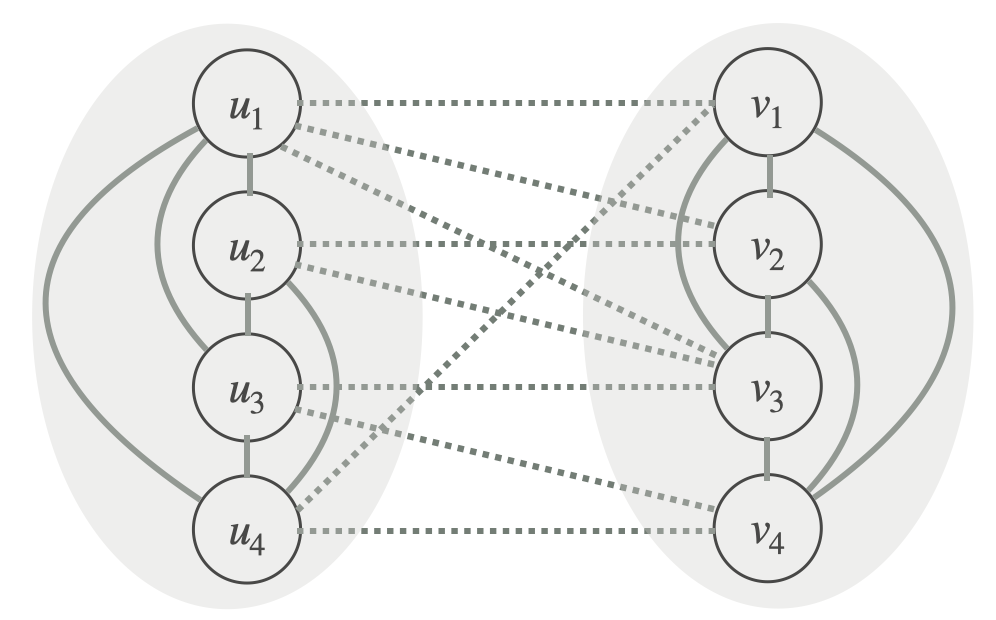}
  \caption{The reason for attemptimg community-merging after all the node-level merging is over. The {\it fp} value of the two communities is ${(6+6)+7 \over 28}={19\over 28}$, and of the merged version is ${6+6+9\over 28}={21\over 28}$. Moving any single vertex from one community into another leads to a reduction in the value of {\it fp}.}
  \label{fig:why-com-merge}
\end{figure}

It starts by assigning each vertex to its own community (line 2). Then it tries to find the best community for each node $u$ to move to, by checking the improvement in the {\it fp} measure after assigning $u$ to each of its neighbouring communities, which are singletons initially (lines 8 -- 27). If such a best community is found, $u$ is then moved to it (lines 23 -- 26). The {\it change} flag is kept to track any changes to the community structure, i.e., set of communities. When there are no changes, then {\it change} equals zero and the algorithms breaks out of the loop and terminates. After all nodes level merges are done, the algorithm attempts to merge whole communities (lines 32 --49). As in the node-level case, the best community to merge with (if it exists) is selected (lines 37 -- 43), and the communities are merged (lines 44--48).   

\begin{figure}[h]
  \centering
  \includegraphics[width=0.3\linewidth]{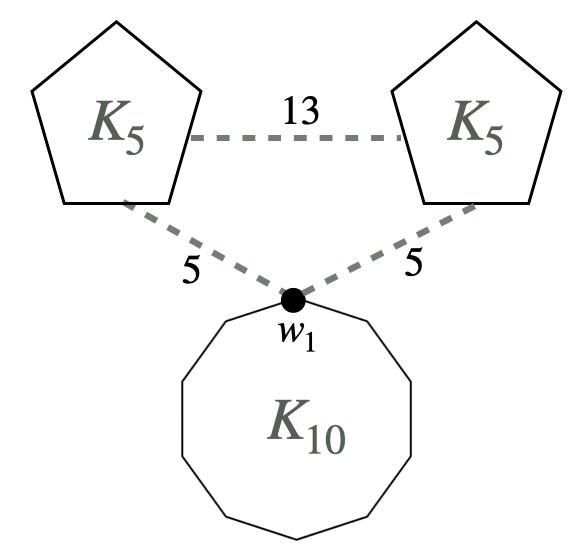}
  \caption{The reason for attempting node-level merging after community merging is done. Node $w_1$ is connected to every vertex in the graph: it is part of a clique of size 10 ($K_{10})$ has 5 neighbours each in the other two communities. The two $K_5$ communities have 13 edges between them (out of a maximum possible value of 25). After the two $K_5$ communities merge, $w_1$ has 10 neighbours in the newly merged community (compared to the 9 it has in its current community). When $w_1$ moves to the merged community above, the {\it fp} value increases from 0.884 to 0.889. Before the two $K_5$ communities merge, moving $w_1$ to any one of the $K_5$ communities reduces the {\it fp} value.}
  \label{fig:why-node-after-com}
\end{figure}

Figure~\ref{fig:why-com-merge} shows an example that shows how community merging could improve the {\it fp} value after all node-level merges are exhausted. Node $u_1$ has 3 neighbours within the community and 3 in the other community, so it has no incentive to switch to the other community. The same is true for $v_3$. Each of the rest of the vertices have more neighbours inside than outside their community. The {\it fp} value of the two communities is ${(6+6)+7 \over 28}={19\over 28}$, and of the merged version is ${6+6+9\over 28}={21\over 28}$.

Figure~\ref{fig:why-node-after-com} shows an example that shows how node-level merging could improve the {\it fp} value after all community level merging is done. Node $w_2$ has 2 neighbours each in the other two communities. After the above two communities merge, $w_2$ has 4 neighbours in the merged community (it has only 3 in its current community). So, it has an incentive to switch to the newly merged community.

\begin{figure}[h]
  \centering
  \includegraphics[width=0.6\linewidth]{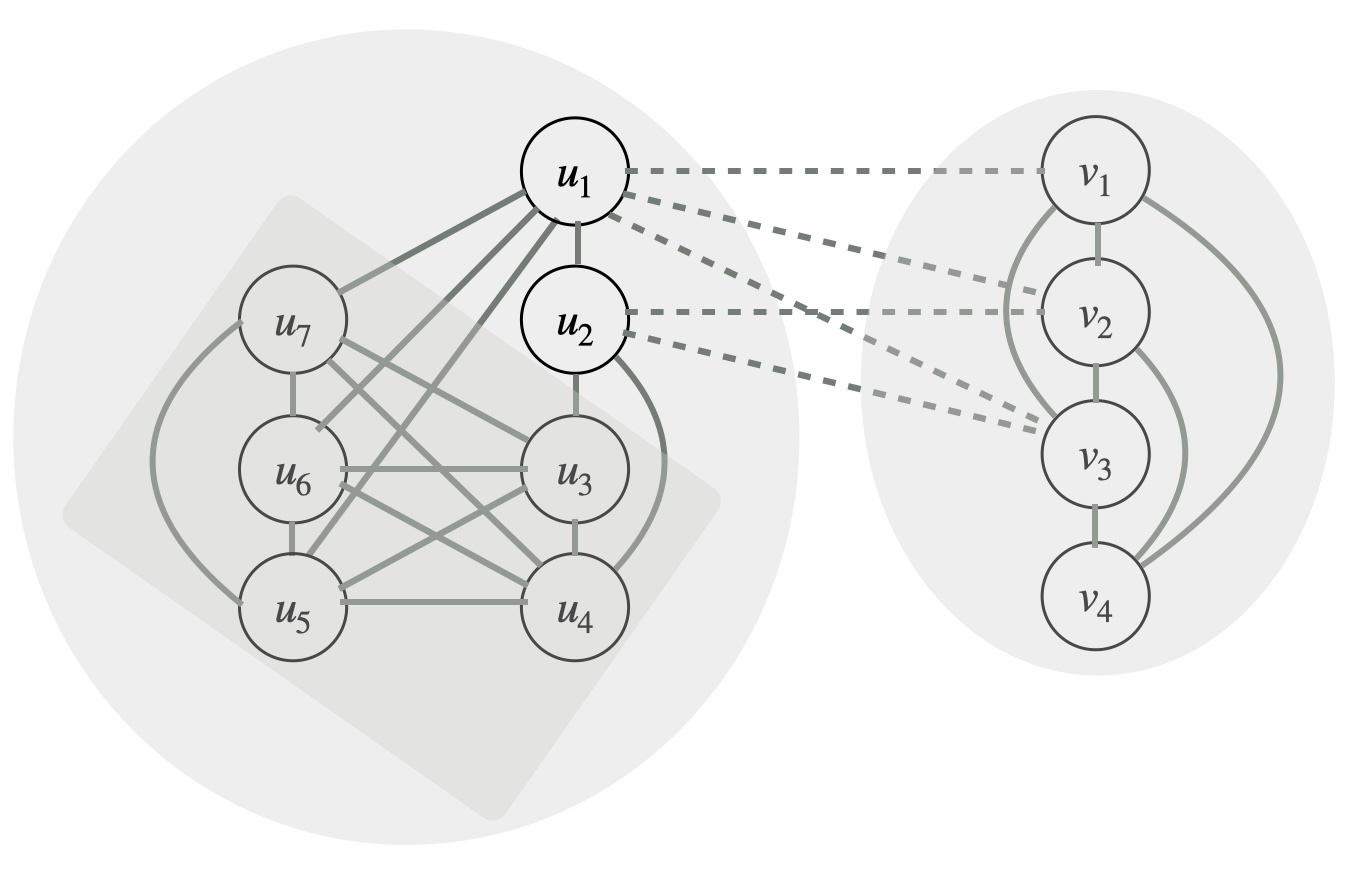}
  \caption{This example shows why Algorithm {\it fpGreed} cannot guarantee an optimal answer. In the above example, the {\it fp} value cannot be improved by considering node-level movements and community merges. However, the {\it fp} value can be improved if we move $u_1$ and $u_2$ together into the community comprising $v_1,v_2,v_3$ and $v_4$.} 
  \label{fig:why-not-optimal}
\end{figure}

{\it Suboptimality of fpGreed.} The algorithm makes a suboptimal choice if $u$ merges with $v$ before $v$ merges with $w$ (if merging the latter first gives a better {\it fp}), so a randomised version of the above may yield better answers. The same problem can occur if $C_i$ merges with $C_j$ before $C_j$ merges with $C_k$ which could have yielded a better {\it fp} value. Another source of sub-optimality is shown in Figure~\ref{fig:why-not-optimal}. It is prohibitively expensive to consider all combinations of vertices for merging.

\begin{table}[h!]
\centering
\caption{Communities in a few Real-World Datasets}
\label{tab:real_world_results_optimization}
\begin{tabular}{lcc}
\toprule
\textbf{Dataset} & \textbf{Performance} & \textbf{No. of Modules} \\
\midrule
Karate Club Network       & 0.9090 & 19 \\
Dolphin Network           & 0.9476 & 28 \\
Florentine Families       & 0.8952 & 7  \\
Les Misérables            & 0.9638 & 34 \\
Football Network          & 0.9583 & 16 \\
\bottomrule
\end{tabular}
\end{table}

Table~\ref{tab:real_world_results_optimization} shows the output of running {\it fpGreed} on a few well-known datasets. This approach is able to identify smaller communities as well. 

\section{A Faster Heuristic}
\label{sec:fast}

Since Algorithm {\it fpGreed} does not scale to large graphs, we propose a faster heuristic in Algorithm {\it fastFP}. {\it fastFP} constructs a weighted graph from the given graph, where the weight on the edge signifies the number of common neighbours between the endpoints and the connections among these common neighbours.
\begin{algorithm}
\caption{{\it fastFp}}
\begin{algorithmic}[1]

\Require Graph $G = (V, E)$
\Ensure A set of communities $\mathcal{C} = \{C_1, C_2, \dots, C_k\}$

\State // Step 1: Compute weight between all node pairs
\Statex // \hspace{.36 in} based on the edges between their common neighbours
\ForAll{$(u, v) \in V \times V, u \ne v$}
    \State $CN \gets$ Common neighbours of $u$ and $v$
    \State $k \gets |CN|$
    \State $e \gets 0$
    \ForAll{$(w_1, w_2) \in CN \times CN, w_1 \ne w_2$}
        \If{$(w_1, w_2) \in E$}
            \State $e \gets e + 1$
        \EndIf
    \EndFor
    \If{$(u, v) \in E$}
        \State $w(u, v) \gets 2k + e + 1$
    \Else
        \State $w(u, v) \gets 2k + e$
    \EndIf
\EndFor

\State // Step 2: Construct a weighted graph $G_2 = (V_2, E_2, W_2)$ 
\Statex // \hspace{.35 in} based on weight threshold $t$
\State $V_2 \gets V$, $E_2 \gets \emptyset, W_2 \gets \emptyset$
\ForAll{$(u, v) \in V \times V, u \ne v$}
    \If{$w(u, v) \ge t$}
        \State Add edge $(u, v)$ with weight $w(u,v)$ to $E_2$
    \EndIf
\EndFor
\State $G_2 \gets (V_2, E_2, W_2)$

\State{// Step 3: Initial Community Detection}
\State $\mathcal{C} \gets []$
\While{$E_2 \ne \emptyset$}
    \State Find edge $(u, v)$ with highest $w(u, v)$
    \State $CN \gets$ Common neighbours of $u$ and $v$
    \State $Community \gets \{u, v\} \cup CN$
    \State Add $Community$ to $\mathcal{C}$
    \State Remove all nodes in $Community$ from $G_2$
\EndWhile

\State{// Step 4: Merge Communities}
\State $Merged \gets$ \textbf{true}
\While{$Merged$}
    \State $Merged \gets$ \textbf{false}
    \ForAll{pairs $(C_1, C_2) \in \mathcal{C}$}
        \State $crossEdgeCount \gets 0$ 
        \ForAll{$u \in C_1$, $v \in C_2$}
            \If{$(u, v) \in E$} 
                \State $crossEdgeCount \gets crossEdgeCount + 1$
            \EndIf
            \State{// a cross-community edge exists only if they are 
            \Statex \hspace{0.55 in}//  connected in the original graph}
        \EndFor
        \If{$crossEdgeCount > \frac{|C_1| \times |C_2|}{2}$}
            \State {// This check ensures that the {\it fp} value improves 
            \Statex \hspace{0.54 in } // upon merging}
            \State Merge $C_1$ and $C_2$ and update $\mathcal{C}$
            \State $Merged \gets$ \textbf{true}
            \State \textbf{break}
        \EndIf
    \EndFor
\EndWhile
\State \Return $\mathcal{C}$
\end{algorithmic}
\end{algorithm}

\begin{table}[h!]
\centering
\caption{Comparison of {\it fpGreed} and {\it fastFp} on Various Datasets}
\label{tab:full_results_table}
\begin{tabular}{|l|c|c|c|c|c|c|}
\hline
Dataset & Nodes & Edges & {\it fpGreed} & {\it fastFp} & {\it fpGreed} &  {\it fastFp}  \\
        &       &       &    {\it fp}   & {\it fp}     & time          &  time   \\
\hline
Karate Club & 34 & 78  & 0.9090 & 0.6791 & $100.2\ ms $ & $83.7\ ms $ \\
Dolphin     & 62 & 159 & 0.9476 & 0.8895 & $177\ ms $   & $90.6\ ms $ \\
Florentine Families & 15 & 20 & 0.8952 & 0.8762 & $86.6\ ms $ & $85.6\ ms$ \\
Les Misérables & 77 & 254 & 0.9648 & 0.8516 & $221.7\ ms$ & $88.1\ ms $ \\
Football Network & 115  & 613      & 0.9583 & 0.8828 & $560.6\ ms $ & $102.7\ ms $ \\
Email EU Core & 1005    & 16706    & 0.9759 & 0.4125 & $1145.839\ ms $ & $1.723\ s $\\
Facebook      & 4039    & 88234    & 0.9933 & 0.9437 & $33695.973\ ms$ & $30.277\ s $ \\
Arxiv dataset & 5242    & 14496    & 0.9995 & 0.9983 & $31495.057\ ms$ & $1.253\ s $ \\
Wikivote      & 7115    & 100762   & -- & 0.7519 & -- & $26.065\ s $ \\
CA-condmat    & 23133   & 93497    & -- & 0.9979 & -- & $19.480\ s $ \\
Internet      & 26475   & 53381    & -- & 0.9678 & -- & $44.009\ s $ \\
Epinions      & 75879   & 405740   & -- & 0.9804 & -- & $336.667\ s $ \\
Slashdot0902  & 82168   & 546487   & -- & 0.9856 & -- & $416.937\ s $ \\
DBLP          & 317080  & 1049866  & -- & 0.9999 & -- & $3935.676\ s $ \\
Web-nd.edu    & 325729  & 1497134  & -- & 0.9987 & -- & $2568.600\ s $  \\
YouTube       & 2987624 & 1134890  & -- & 0.9937 & -- & $41476.909\ s $ \\
Live Journal  & 3997962 & 34681189 & -- & 0.9997 & -- & $573574.157\ s $  \\
\hline
\end{tabular}
\end{table}

The first step is to create a weighted graph $G_2$ based on the neighbourhood of all pairs of nodes  that may or may not be connected by an edge in the input graph $G$ (lines 2 -- 16). This step also creates an opportunity for a node pair $u, v$ to belong to the same community if they have several common neighbours even if an edge does not exist between them. The weight depends on the edges among the common neighbours of $u$ and $v$ (lines 6 --10). The next step is to create a weighted graph in which two vertices $u$ and $v$ may be connected by a weighted edge if the weight exceeds a threshold $t$. We set $t = 3$ (the smallest possible value) for all the experiments in this paper. Choosing a larger value should reduce the running time at the expense of the {\it fp} value. A weighted edge may be created between $u$ and $v$ even if they were not connected by an edge in $G$. Step 3 (lines 26 -- 33) create an initial set of communities by considering edges in non-increasing order of edge weight. This is done to ensure that the strongest community is formed first. A node $w$ may be a common neighbour to
two (or more) node pairs. This step ensures that it joins the strongest pair and is then removed from consideration for other communities (line 32). The next step (lines 35 -- 53) is to merge communities if the {\it fp} value improves (see Lemma~\ref{lem-mrg} for details). 


\begin{lemma}
\label{lem-mrg}
Consider two communities $C_1$ and $C_2$ with $m$ and $n$ vertices respectively. Let $e_m$ and $e_n$ denote the number of internal edges within $C_1$ and $C_2$ respectively. If the number of cross-edges (inter-community edges) $e_{cc}$ between the two communities is greater than $m\times n\over 2$, then the {\it fp} value after merging them is higher. Before merger,
\begin{eqnarray}
{\it fp}{\rm (separate)} &=& {| e_m + e_n | + | (m\times n) - e_{cc} |\over m\times n}\\
{\it fp}{\rm (merged)} &=& {| e_m + e_n + e_{cc} |\over m\times n}\\
{\rm if\ } e_{cc} &>& {m\times n\over 2}, \\
{\it fp}{\rm (merged)} &>& {\it fp}{\rm (separate)}.
\end{eqnarray}
\end{lemma}

\section{Experiments}
\label{sec:expt}

Table~\ref{tab:full_results_table} shows the {\it fp} values and the time(s) taken to obtain them by running Algorithms {\it fpGreed} and {\it fastFp} on various datasets. Expectedly, {\it fastFP} runs faster than {\it fpGreed}, but may yield poorer {\it fp} values. The experiments were run on a Mac Studio Apple M2 Max with 12-core CPU, 38-core GPU, 16-core Neural Engine, 96GB unified memory and 4TB SSD storage. The algorithms are implemented in Python~\cite{python_ref} using Networkx~\cite{networkx}. 
The timings were calculated using the command-line benchmarking tool hyperfine~\cite{Peter_hyperfine_2023}. For Algorithm {\it fastFP}, we set the threshold parameter $t =3.$ We could do multiple runs of the algorithms on the smallers datasets only. 

For the larger datasets in the table, {\it fastFp} appears to yield good results, both in terms of the quality of the output as well as running time. 

\section{Discussion and Future Work}
\label{sec:conclude}
Optimising the {\it fp} measure appears to be a promising direction to pursue for community detection based on the preliminary results obtained in this paper. The following important questions need to be addressed: 
\begin{enumerate}
\item {\it Limitation: Null models.} Modularity compares the number of links inside a given module
with the expected value for a randomized graph of the same size
and same degree sequence~\cite{fortunato2007resolution}. But it suffers from the resolution limit problem. On the other hand, an optimal {\it fp} value would yield the best possible communities based on an intuitively agreed upon notion of communities in a network, i.e., the number of edges within communities should be as large as possible and the number of edges across communities should be as small as possible. And it is parameter-free. But the {\it fp} measure formulation does not account for null models. Is there a way to separate null model considerations from the parameters we optimise for community detection? For example, does it make sense to divide the network into communities by optimising {\it fp}, and then use this community structure to measure the modularity? 
\item {\it Limitation: Unweighted graphs.} The current formulation is for unweighted graphs only. Can the {\it fp} measure be generalised to weighted graphs? One possibility is to consider the non-edges to be of weight 1, leading to the following:
    \begin{equation}
{\textit fp_{wt}}({\mathcal P}) = { \sum_{(i,j)\in E, C_i = C_j} w(i,j)  + |\{(i,j)\not\in E, C_i\not= C_j\} |\over {\sum_{i,j\in E} w(i,j)}+|\{(i,j)\not\in E, C_i\not= C_j|}
    \end{equation}
    However, if the edges in the graph have high weights, this may lead to a bias towards adding all the edges to a single community. 
\item {\it Future work: Speed.}\\
(a) How to further speed up both {\it fpGreed} and {\it fastFp} is another question of interest. For example, the idea from Lemma~\ref{lem-mrg} can be used in {\it fpGreed} to cut down some more unnecessary computations. \\
(b) The role of the threshold $t$ in {\it fastFp} needs to be explored further. Currently, it is set to the minimum possible value. Raising it ought to improve the speed, but may reduce the quality of the solution. 
\item {\it Future work: Exploring the effects of optimising {\it fp}.} The distribution of the community sizes and number of communities obtained by these methods needs to be explored further.
\end{enumerate}
\section*{Acknowledgments}
This work was supported in part by a grant from Cisco Research, San Jose, USA.
\bibliographystyle{unsrt}  
\bibliography{main-base}

\end{document}